\documentclass[aip,twocolumn,reprint,groupedaddress]{revtex4-2}
\usepackage{amsmath}
\usepackage{graphicx}
\usepackage{dcolumn}
\usepackage{bm}
\usepackage{siunitx}
\usepackage{xcolor}

\begin{document}


\title{Collision efficiency of droplets across diffusive, electrostatic and inertial regimes}


\author{F. Poydenot, B. Andreotti}
\affiliation{Laboratoire de Physique de l'Ecole Normale Sup\'erieure (LPENS), CNRS UMR 8023, Ecole Normale Sup\'erieure, Universit\'e PSL, Sorbonne Universit\'e, and Universit\'e de Paris, 75005 Paris, France}


\date{\today}

\begin{abstract}
Rain drops form in clouds by collision of submillimetric droplets falling under gravity: larger drops fall faster than smaller ones and collect them on their path. The puzzling stability of fogs and non-precipitating  warm clouds with respect to this avalanching mechanism has been a longstanding problem. How to explain that droplets of diameter around $10~{\rm \mu m}$ have a low probability of collision, inhibiting the cascade towards larger and larger drops? Here we review the dynamical mechanisms that have been proposed in the literature and quantitatively investigate the frequency of drop collisions induced by Brownian diffusion, electrostatics and gravity, using an open-source Monte-Carlo code that takes all of them into account. Inertia dominates over aerodynamic forces for large drops, when the Stokes number is larger than $1$. Thermal diffusion dominates over aerodynamic forces for small drops, when the Péclet number is smaller than $1$. We show that there exists a range of size (typically $1-10~{\rm \mu m}$ for water drops in air) for which neither inertia nor Brownian diffusion are significant, leading to a gap in the collision rate. The effect is particularly important, due to the lubrication film forming between the drops immediately before collision, and secondarily to the long-range aerodynamic interaction. Two different mechanisms regularise the divergence of the lubrication force at vanishing gap: the transition to a noncontinuum regime in the lubrication film, when the gap is comparable to the mean free path of air, and the induction of a flow inside the drops due to shear at their surfaces. In the gap between inertia-dominated and diffusion-dominated regimes, dipole-dipole electrostatic interactions becomes the major effect controlling the efficiency of drop collisions.
\end{abstract}


\maketitle
\begin{figure*}[t!]
\centering
\includegraphics{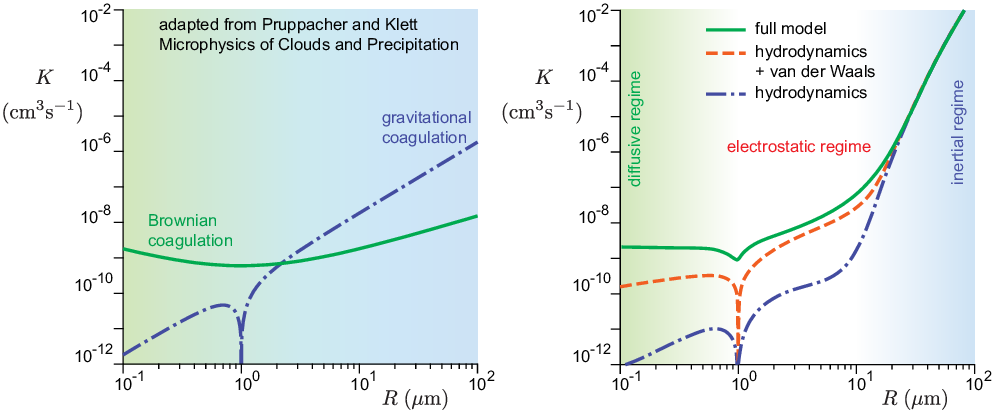}
\caption{A change of conceptual model. (a) Graph adapted from the textbook of \citeauthor{pruppacher_microphysics_2010} \citep[chap.~15]{pruppacher_microphysics_2010}, showing the current vision in cloud microphysics. Comparison between two collisions modes for a spherical particle of $1\;{\rm \mu m}$ interacting with a second particle of radius $R$, showing a cross-over between Brownian coagulation (green solid line) and gravitational coagulation (dotted-dashed blue line). The calculations are based on \citeauthor{klett_class_1975}\cite{klett_class_1975} and \citeauthor{hidy_removal_1973}\cite{hidy_removal_1973}. (b) Predictions made here, including gravity, inertia and hydrodynamics (dotted-dashed blue line), adding van der Waals interactions (dashed red line) and then, Brownian diffusion (green solid line). A new regime appears between the diffusive regime and the inertial regime, where electrostatic effects become dominant.
\label{fig:ChangeParadigm}
}
\end{figure*}

\section{Introduction}
\subsection{Cloud microphysics and collisional aggregation of droplets}
Collisional aggregation of water droplets is at the core of cloud microphysics.\cite{pruppacher_microphysics_2010} Current atmospheric global circulation models, used for climate modeling, are based on phenomenological formulations for the evolution equation of drop populations.\cite{cotton_storm_2011,hansen_global_2023,schmidt_ceresmip_2023} The drop population is represented by a few moments of its distribution, with empirically determined rate coefficients for each process.\cite{kessler_distribution_1969,morrison_confronting_2020} This has mainly the advantage of low computational cost. More sophisticated techniques involve solving the distribution over size bins in an Eulerian description,\cite{khain_notes_2000-1,khain_representation_2015} or simulating a small number of representative "super-droplets" in a Lagrangian description.\cite{shima_super-droplet_2009,grabowski_modeling_2019} However, in all cases the aggregation coefficients must still be computed \textit{a priori} to accurately describe the microphysics at play at the population level. Efficiencies reported in the literature are sometimes inconsistent and computed over narrow ranges of sizes so that the crossover regimes between different mechanisms are still poorly resolved.\cite{khain_notes_2000-1}

In warm clouds, i.e. in the absence of ice crystals, drops nucleate on hydrophilic aerosol particles, named cloud condensation nuclei. The scavenging and removal from the atmosphere of micronic and submicronic particulate matter by millimetric raindrops has been widely studied since the 1957 work of Greenfield,\cite{greenfield_rain_1957} particularly in the context of atmospheric pollution.\citep{ervens_modeling_2015} For particle sizes between $0.1\,{\rm \mu m}$ and $2.5\,{\rm \mu m}$, a range often called the "Greenfield gap", the scavenging of pollutants by raindrops is inefficient and particles can stay suspended in the atmosphere for very long times, from weeks to months.\citep{friedlander_smoke_2000} When humid air rises by convection, its relative humidity increases until the lifting condensation level is reached and droplets nucleate. Condensation growth stops when the humidity in the air between droplets approaches saturation, from supersaturated values.\citep{twomey_nuclei_1959,ghan_droplet_2011-1} The volume fraction of liquid water in clouds is controlled thermodynamically by the liquid-vapor coexistence curve and is typically lower than $10^{-6}$. This constrains the tradeoff between the typical drop size and the number of drops per unit volume: the cloud condensation nuclei density selects a large number of small drops, rather than a small number of large drops \citep{krueger_technical_2020}. The number of drops per unit volume in warm clouds is typically\citep{hess_optical_1998} $\psi \sim 10^8\;{\rm m^{-3}}$ so that condensation growth leads to micron scale droplets. The concentration of raindrops in clouds is typically $10^{-5}$ smaller than the concentration in micron-size droplets. In order to grow from $10\;{\rm \mu m}$ (cloud-drop) to $1\;{\rm mm}$ (raindrop), a drop would have to pump the water content of $300\;{\rm cm^3}$ of droplet-free air: this is totally inconsistent with observations, as this volume typically contains $30000$ drops. Rain in warm clouds must therefore form by collision and coalescence of cloud droplets\citep{beard_warm-rain_1993,mcfarquhar_rainfall_2022} — one million droplets of $10\;{\rm \mu m}$ radius are needed to form a millimeter size raindrop.

The collisional behaviour of droplets near the size range of $0.1-10\,{\rm \mu m}$ is still poorly understood, especially for small drops of commensurable sizes. Notably, the drop size distribution in clouds is observed to broaden over time as droplets grow,\citep{brenguier_droplet_2001} which would involve initially the interaction of micronic droplets of similar sizes. The scientific literature reveals an open problem in understanding the stability of mists and clouds with respect to the aggregation of their liquid water into drizzle and rain. Why do some warm clouds remain stable for long periods of time while others form precipitation? Why are mists and fogs stable? On the one hand, the growth of droplets to the micron scale can be explained by the individual condensation growth of each drop, without any collective effect. On the other hand, the growth of raindrops by accretion of smaller drops during their fall under the effect of gravity explains the precipitation phenomenon. But how to explain that the growth of raindrops by coalescence is inhibited in mists and clouds? How to explain symmetrically that growth occurs over the $1-10\,{\rm \mu m}$ gap in precipitating clouds? The aim of this paper is to shed light on this issue using a detailed model of collision frequency which combines all the effects discussed in the literature.

A large part of this paper is devoted to the description and to the analysis of the model, which reviews the dynamical mechanisms that have been previously included in the investigation of drop collision efficiency. The originality of the paper results from the combination of the effects of Brownian diffusion, electrostatics,  hydrodynamics and inertia, which allows us to compare them and unravel the existence of a range of drop size for which electrostatics effect are dominant.
\begin{figure*}[t!]
\centering
\includegraphics{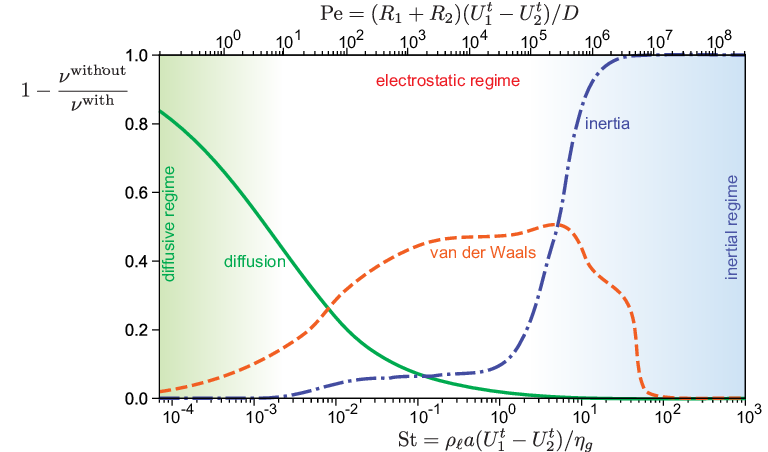}
\caption{Effect of the various dynamical mechanisms on the collision rate for $\Gamma= R_1/R_2 = 5$. Solid green line: difference in the collision rate with and without thermal diffusion, compared to the collision rate with diffusion, as a function of the Péclet number (top axis). Dash-dotted blue line:  difference in the collision rate with and without inertia, compared to the collision rate with inertia, as a function of the Stokes number (bottom axis). Inertia dominates at $\mathrm{St}>1$ while thermal diffusion dominates at $\mathrm{Pe}<1$, leaving $3$ decades in $\mathrm{St}$ where neither inertia nor thermal diffusion are relevant. Dashed orange line: difference in the collision rate with and without van der Waals interactions, compared to the collision rate with van der Waals interactions. Lubrication is the dominant effect to reduce the collision rate between the diffusive and inertial regimes. van der Waals interactions limits this gap, but only accounts for a fraction of it.
\label{fig:WhichMechanism}
}
\end{figure*}

\subsection{Collisional efficiency}
At lowest order, the problem of collisional growth of a drop population can be described only with binary collisions. A collector drop of mass $m_1$ and radius $R_1$, collecting smaller drops inside a homogenous cloud of droplets of mass $m_2$, radius $R_2$ and number concentration $n_2$ grows at a rate
\begin{equation}
\frac{\mathrm d m_1}{\mathrm d t} =  m_2 \nu.
\label{eq:growthrate_general}
\end{equation}
$\nu$ is the collision frequency of drops $1$ and $2$. In this case, $\nu \equiv K n_2 $ is proportional to $n_2$, as more drops means more collisions. $K$ is called the collisional kernel between drops of size $R_1$ and $R_2$. $K$ generally depends on the sizes of the drops through the particular collision mechanism driving them together. In the case of gravitational collisions, both drops fall at their terminal velocities $U_1^t$ and $U_2^t$. The growth rate of the collector drop $1$ is thus
\begin{equation}
\frac{\mathrm d m_1}{\mathrm d t} = m_2 n_2 K \;\;\mathrm{with}\;\; K= \pi (R_1+R_2)^2 \left|U_1^t-U_2^t\right| E.
\label{eq:growthrate}
\end{equation}
$\pi (R_1+R_2)^2 \left|U_1^t-U_2^t\right|$ is the volume swept by unit time as the two drops settle. $E$ is called the collision efficiency. It is the dimensionless collision cross-section induced by aerodynamic interactions: for geometric collisions, $E = 1$. The problem is not a simple two-body problem, but a three-body one: the third body is air. $E$ therefore depends on the drop characteristics, their initial velocities and the flow between the two. $E$ has been measured for water drops in air using three different methods. The first method relies on making a single collector droplet fall in still air into a monodisperse cloud of smaller droplets with dissolved salt inside. The size and salt concentration of collector droplets is measured, which allows a determination of $E$ knowing the properties of the cloud.\citep{picknett_collection_1960,woods_experimental_1964,beard_measurement_1979,beard_measured_1983,ochs_iii_laboratory_1984} The main uncertainties come from determining the droplet cloud properties, and ensuring the collector drop actually falls at its terminal velocity.\citep{chowdhury_free_2016} The second method relies on keeping a collector drop afloat in a wind tunnel by dynamically matching the flow speed to its terminal velocity. The collector drop impacts with a number of smaller drops; measuring the collector terminal velocity allows to determine its mass, therefore its growth rate and the collision efficiency.\citep{gunn_laboratory_1951,beard_wind_1971-1,levin_experimental_1973,abbott_experimental_1974,vohl_collision_2007} Lastly, some authors\citep{schotland_collision_1957,telford_observations_1961,woods_wake_1965,beard_experimental_1968,low_collision_1982} make two individual drops fall by in still air and directly measure their trajectories. All these methods are limited by uncertainties around $10\;\%$, and very little data is available about drops below $30\;{\rm \mu m}$ colliding with drops of similar sizes.

The efficiency for droplets with significant inertia is well captured by most models, as the collision is controlled by the long-range aerodynamic interaction. However, predicting in the overdamped regime if two colliding drops merge is extremely sensitive to the modeling details, where the efficiency reaches a minimum. For instance, in the Stokes approximation, the interaction between two spheres via the lubrication air film increases as the inverse of the gap $H$ between them: collisions cannot happen in a finite time. Several mechanisms regularize this singularity. Shear at the drop surface induces a flow inside the drop, which changes the short-range behaviour of the force and allows collisions in a finite time. At separations comparable with the mean free path $\bar \ell$ of the carrying gas, slip flow between the drops due to the rarefaction of air regularizes the force, bringing it to a weak logarithmic divergence at contact. Van der Waals forces also help bring the drops together. Small enough droplets diffuse, which can couple to all of these effects. Together, all these mechanisms create a gap in the collision rate of micron-scale water droplets where they are all of the same magnitude. This makes the collisional aggregation in this range of drop sizes both difficult to measure experimentally and to model accurately. The model introduced here is both tractable mathematically and exhaustive from the mechanistic point of view to gain an understanding of each microphysical effect on its own.

As we aim here to take into account Brownian diffusion in the same model as gravity, we will extend the concept of collisional efficiency in section \ref{sec:diffusion} by adapting the reference collision frequency.
\begin{figure*}
\centering
\includegraphics{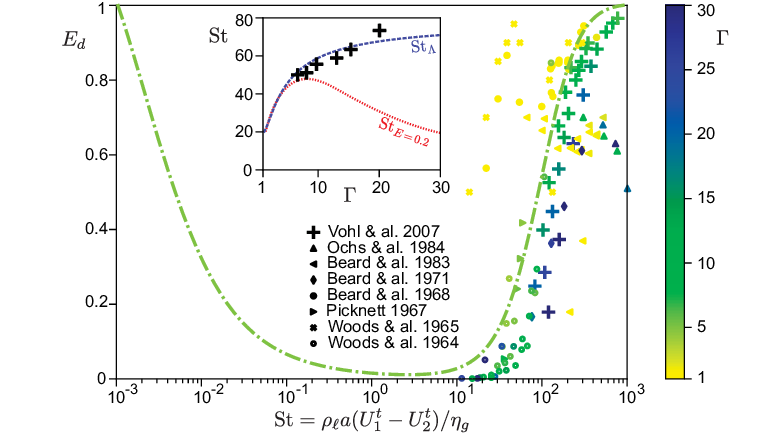}
\caption{Experimental dataset of the collision efficiency corrected for diffusion $E_d$ as a function of the Stokes number $\mathrm{St} = \rho_\ell a (U_1^t-U_2^t)/\eta_g$. Color indicates radius ratio $\Gamma=R_1/R_2$. The dash-dotted green line is the computed curve for $\Gamma=5$. Insert: transitional Stokes number as a function of the radius ratio $\Gamma$. Dashed blue line: critical Stokes number $\mathrm{St}_\Lambda$ for head-on collisions given by Eq.~(\ref{eq:critical_stokes}), with $\Lambda = 4.7$. Dotted red line: Stokes number for the full model at which the efficiency reaches $E=0.2$ in the inertial regime. Crosses: Stokes number for which $E=0.2$ obtained from fitting $\Gamma$-aggregated data from \citeauthor{vohl_collision_2007} \cite{vohl_collision_2007} to the full model presented here. Agreement is good for $\Gamma <10$ but degrades at larger size ratios. The simple head-on collision model~[Eq.~(\ref{eq:critical_stokes})] captures well the observed trend in the experimental data.}
\label{fig:comparisonexp}
\end{figure*}

\subsection{Dynamical mechanisms}
Various techniques have been used to formulate the aerodynamic interactions. In the Stokes approximation, there exists an exact solution formulated in bispherical coordinates over the entire flow domain for two solid spheres moving at equal velocities along their line of center.\cite{stimson_motion_1926} This solution was extended to the case of different velocities,\citep{maude_end_1961} a sphere moving towards a plane,\citep{brenner_slow_1961} and two droplets moving along their line of center.\citep{haber_low_1973} Approximate solutions for more general flow configurations have been investigated using the method of reflections\citep{happel_low_1981,hetsroni_low_1978} and twin multipole expansions.\citep{jeffrey_calculation_1984,jeffrey_calculation_1992} These techniques give solutions as series converging rapidly when the drops are far apart, but requiring an increasingly larger number of terms as the gap vanishes. The interaction can thus be decomposed into a long-range part, due to viscous forces between the drops, and a short-range part, due to lubrication squeeze flow at vanishing gaps. The lubrication force between a sphere and a plane has been computed with a matched asymptotic expansion.\cite{cooley_slow_1969,oneill_asymmetrical_1970} The force due to drop flow has been determined using a boundary-integral formulation \citep{jansons_general_1988} in the limit of non-deformable drops.\cite{davis_lubrication_1989} Drop deformation has been investigated for slightly deformable drops of very different sizes under van der Waals attraction,\cite{yiantsios_close_1991} and numerically at finite capillary number.\cite{zinchenko_novel_1997} Dilute gas effects are of two types. First, when the continuum approximation still holds, gas molecules can bounce along the surface, leading to slip boundary conditions with a slip length close to the mean free path $\bar\ell$. Hocking\cite{hocking_effect_1973} first showed that it leads to collisions in a finite time between a sphere and a plate. These results were extended to the collisions of two spheres.\cite{barnocky_effect_1988} The multipole expansion\cite{jeffrey_calculation_1984} has also been extended to the case of diffusive reflective molecular boundary conditions.\cite{ying_hydrodynamic_1989} When the gap is smaller than $\bar\ell$, the continuum approximation underlying the Navier-Stokes equations itself breaks down, and the full Boltzmann transport equation must be solved. Such free-molecular Poiseuille flow between two parallel planes was solved\cite{cercignani_flow_1963,hickey_plane_1990} using a BGK\citep{bhatnagar_model_1954} approximation of the Boltzmann equation. This solution has been extended to the case of two approaching drops.\cite{sundararajakumar_non-continuum_1996} A uniform approximation between this solution and the multipole expansion\cite{jeffrey_calculation_1984} has also been determined.\cite{li_sing_how_non-continuum_2021}

The collision efficiency $E$ is of very practical interest to cloud physics modeling, as it directly determines the collisional growth rate. $E$ has first been computed by Langmuir,\cite{langmuir_production_1948} who showed that warm clouds, above $\SI{0}{\celsius}$, can produce rain. Following this seminal paper, the basic idea was to compute $E$ using a linear superposition in the Oseen approximation of the flows created by the two individual drops, and assuming near-contact lubrication was negligible.\cite{pearcey_theoretical_1957} Various authors \citep{shafrir_collision_1963,klett_theoretical_1973,schlamp_numerical_1976,pinsky_collision_2001} improved upon this formulation by using more accurate formulations at finite Reynolds numbers of the flow around a single drop. Hocking\cite{hocking_collision_1959} used instead a linear superposition of Stokes solutions, and predicted that there was a critical size below which no collisions would occur. Linear superposition does not naturally verify the correct boundary conditions at the drop surfaces. However, no slip boundary conditions can be verified on angular average around the drop,\cite{wang_improved_2005} but all superposition methods fail to reproduce the divergent force behaviour at vanishing gaps. To correctly capture this, it has been proposed to decompose the aerodynamic interaction into a divergent short-range force and a long-range force computed using the superposition method.\cite{rosa_accurate_2011} All the superposition schemes without short-range interactions detect collisions using arbitrary distance thresholds below which contact is said to occur. Davis\cite{davis_theoretical_1967} and Hocking\cite{hocking_collision_1970} used formulations of the force based on the Stokes solution in bispherical coordinates,\cite{stimson_motion_1926} with an arbitrary cutoff distance; the results for drops below $\SI{20}{\mu m}$ were particularly sensitive to the value chosen. Slip flow was introduced later\cite{davis_collisions_1972,jonas_collision_1972} to the Stokes solution\cite{stimson_motion_1926} and the need of an arbitrary cutoff was removed. Recently, \citeauthor{ababaei_collision_2023}\cite{ababaei_collision_2023} compared in Stokes flow the twin multipole expansion with an analytical solution in bispherical coordinates and noncontinuum lubrication.\cite{reed_particle_1974} \citeauthor{rother_gravitational_2022}\cite{rother_gravitational_2022} also made use of bispherical coordinates, considering flow inside the drops, slip as the only noncontinuum effect and the effect of van der Waals forces. It must be noted that the drop Reynolds number reaches $1$ around a particle radius of $\SI{56}{\mu m}$, making the applicable range of Stokesian aerodynamics very limited in this problem.\cite{guazzelli_physical_2012}

The effect of Brownian diffusion on gravitational collisions has been studied through the lens of small particle-droplet interactions. These effects are often taken to be additive,\citep{greenfield_rain_1957,slinn_approximations_1977} yielding approximate collision rates that cannot reflect coupling between these mechanisms. The problem of mass transport to a sphere, thus neglecting particle inertia, has been investigated by solving a diffusion-advection problem with a given flow around the large drop. Matched asymptotic expansion gives an approximate solution for Stokes flow.\cite{friedlander_mass_1957,acrivos_heat_1962} An analytical collision rate for two droplets ignoring all aerodynamic interactions has also been derived.\cite{simons_kernel_1986} The collision rate for non-inertial droplets in Stokes flow has been computed with a Fokker-Planck equation for the pair distribution function, including van der Waals forces.\cite{zinchenko_gravity-induced_1994,zinchenko_collision_1995} Correctly handling particle inertia can only be done by integrating a Langevin equation for the problem. The collision efficiency for particles of very different sizes was computed using Monte-Carlo simulations taking into account aerodynamic interactions without short-range lubrication, particle inertia, but also electrostatic forces, thermophoresis and diffusiophoresis.\cite{tinsley_electric_2010,tinsley_charge_2013,tinsley_parameterization_2015,zhang_parameterization_2018,cherrier_aerosol_2017,depee_theoretical_2019} Electrostatic effects due to static fields or droplet charges were shown both theoretically and experimentally to lead to enhanced collision rates,\cite{sartor_electrostatic_1960,sartor_role_1967,hocking_collision_1970,ochs_charge_1987,zhang_theoretical_1995,grashchenkov_interaction_2011,magnusson_collisions_2022} with unclear consequences on cloud physics. van der Waals interaction was later taken into account,\cite{yiantsios_close_1991,rosa_accurate_2011,rother_buoyancy-driven_1997,rother_gravitational_2022} without considering air inertia and thermal diffusion.

To the best of our knowledge, there are no computations or measurements of the efficiency for two water droplets, in air, considering at once droplet inertia, inertial effects in the gas flow, noncontinuum lubrication, flow inside the drops, Brownian motion and van der Waals interactions for drops of all relative sizes over the whole $0.1-100\;{\rm \mu m}$ size range most relevant to the rain formation process and the stability of fogs and clouds.
\begin{figure*}[t!]
\centering
\includegraphics{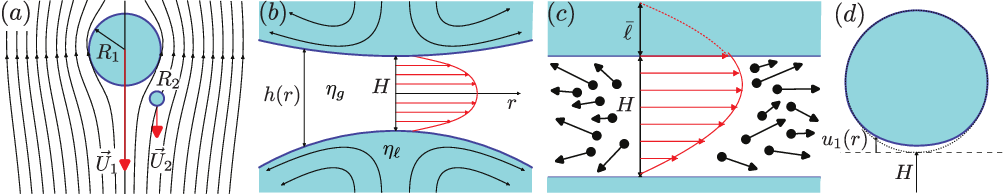}
\caption{Aerodynamic mechanisms involved in drop collisions. (a) long-range aerodynamic interaction. (b) Lubrication in the air film near contact. The shear of the squeezing air flow creates a flow inside the drops. (c) When the gap is comparable to the mean free path $\bar \ell$, rarefaction of the air between the drops leads to partial slip boundary conditions and a lower effective viscosity. (d) The pressure inside the lubrication air film induces a capillary flattening of the interface.}
\label{fig:schemaforces}
\end{figure*}

\subsection{Organisation of the paper}
In this article, we compute the collision efficiency of two settling water drops in air. In section~\ref{sec:dynamicalregimes}, we analyse the dimensionless numbers controlling the three regimes (inertial, electrostatic and diffusive) and summarise our findings. We review experimental data available in the literature, compare them to our calculations and highlight the parameter range in which mechanistic knowledge is lacking. Then, we detail the three regimes. We consider the athermal limit of the problem in section~\ref{sec:inertial} and analyse the transition from inertial to electostatic regimes. We decompose the aerodynamic interaction into two parts: a long-range contribution due to the viscous disturbance flow created by the drops, and a short-range contribution due to the squeezing flow pressure between the drops near contact. We combine the results of \citeauthor{davis_lubrication_1989}\cite{davis_lubrication_1989} and \citeauthor{sundararajakumar_non-continuum_1996}\cite{sundararajakumar_non-continuum_1996} with the well-known lubrication theory into a single analytical, uniformly valid formula. We interpret the results at the light of the different physical mechanisms involved, and explain the behaviour of the collisional efficiency using analytic results for head-on frontal collisions between drops. van der Waals interactions are added in section \ref{sec:electrostatic}, where the electrostatic dominated regime is discussed. Finally, the unification of gravitational, electrostatics and Brownian coagulation is considered in section~\ref{sec:diffusion}. Starting from the collision frequency, we define a combined diffusiogravitational efficiency, to serve as a reference case when computing collision rates with different mechanisms. We compute this new efficiency using Monte-Carlo simulations, and discuss the additivity of gravitational and Brownian coagulation modes.

\section{Dynamical regimes}
\label{sec:dynamicalregimes}

\subsection{Dimensionless numbers}
We consider two liquid drops noted $1$ and $2$ falling under gravity in a gas and subject to thermal diffusion. The position of their center of mass is noted $\vec r_i$ and their radii, $R_i$. The first dimensionless number in the problem is the drop radius ratio %
\begin{equation}
\Gamma=\frac{R_1}{R_2}\ge 1.
\end{equation}
The curvature of the gap between the drops depends on the characteristic drop size: 
\begin{equation}
a \equiv R_1R_2/(R_1+R_2).
\end{equation}
$a$ varies from $R_2/2$ when both drops have the same radius and $R_2$ when $R_1$ is much larger than $R_2$. The gas mean free path $\bar \ell$ plays an important role in the problem, as it controls the transition to the noncontinuum Knudsen aerodynamical regime. A second dimensionless number is therefore
\begin{equation}
{\mathcal A}\equiv\frac{a}{\bar \ell} = \frac{R_1R_2}{(R_1+R_2)\bar \ell}.
\end{equation}

Two further parameters compare the viscosity of the liquid $\eta_\ell$ and that of the gas $\eta_g$, and the density of the liquid $\rho_\ell$ and that of the gas $\rho_g$:
\begin{equation}
{\mathcal N} \equiv \frac{\eta_\ell}{\eta_g}\quad {\rm and } \quad \mathcal D \equiv \frac{\rho_\ell}{\rho_g}.
\end{equation}
We will consider water drops in air at $\SI{25}{\celsius}$ and $\SI{1}{atm}$ for which $\eta_g=18.5\times10^{-6} \,{\rm Pa.s}$, $\eta_\ell=8.9\times10^{-4} \,{\rm Pa.s}$, $\rho_g=1.2\,{\rm kg.m^{-3}}$, $\rho_\ell=1000\,{\rm kg.m^{-3}}$ and\citep{jennings_mean_1988} $\bar \ell=68\;{\rm nm}$. The dimensionless numbers are therefore ${\mathcal N}=48$ and ${\mathcal D}=830$. The influence of inertia in the drop dynamics is controlled by the dimensionless number
\begin{equation}
{\mathcal G}=\left( \frac{ \rho_\ell (\rho_\ell-\rho_g) g } {\eta_g^2} \right)^{1/3} \;a.
\end{equation}
The ratio ${\mathcal G}/{\mathcal A}$ does not depend on the drop sizes and is equal to $2\times 10^{-2}$ for water drops in air. We introduce the typical radius $b$ at which ${\mathcal G}$ is equal to $1$:
\begin{equation}
b=\left( \frac{\eta_g^2}{ \rho_\ell (\rho_\ell-\rho_g) g } \right)^{1/3}.
\end{equation}
For water drops in air, we get $b = 3.3\;{\rm \mu m}$, which is the typical size of drops in clouds and fogs. The thermal noise is controlled by the dimensionless number
\begin{equation}
\mathcal{K} = \frac{3\rho_\ell kT}{4\pi\eta_g^2 \bar\ell}.
\end{equation}
At ambient temperature, for water, it is around $\mathcal{K} = 4.2\,10^{-2}$. The surface tension $\gamma$ controls both drop deformations and van der Waals interaction. It gives the dimensionless number
\begin{equation}
{\mathcal S}=\frac{\rho_\ell \gamma }{\eta_g^2}\;a.
\end{equation}
The ratio ${\mathcal S}/{\mathcal A}$ does not depend on the drop sizes and is equal to $1.4 \times 10^4$. Two electrostatic effects are taken into account: van der Waals interactions are characterised by surface tension and by the Hamaker constant $H$, which can be made dimensionless using
\begin{equation}
{\mathcal T}=\frac{\sqrt{H/(24\pi \gamma)} }{\bar \ell}.
\end{equation}
${\mathcal T}$ is equal to $1.21 \times 10^{-3}$ for water.

\subsection{Dynamical equations}
We consider that both drops are entrained by the same background fluid velocity and denote by $\vec V_i$ their velocity with respect to this background velocity. The aerodynamic interaction is decomposed into a long-range contribution due to viscous stress, computed using the Oseen approximation and a short-range contribution due to pressure, computed in the lubrication approximation, as shown in Fig.~\ref{fig:schemaforces}. We neglect the gradients of velocity at the scale of $R_1$ and $R_2$. We denote by $H=|\vec r_2-\vec r_1|-R_1-R_2$ the distance between drops. We will consider the Oseen approximation, valid at Reynolds number ${\mathcal R}_i \ll 1$. The equation of motion of drop $i$ reads
\begin{equation}
\begin{split}
\frac 43 \pi R_i^3 \rho_\ell \frac{d \vec V_i}{dt} = \frac 43 \pi R_i^3 (\rho_\ell-\rho_g) \vec g - 6 \pi \eta_g R_i\left(1+\frac 38 {\mathcal R}_i \right) {\vec V_i} \\
+\vec F_{ji} -6 \pi \eta_g a^2 \zeta'(H) \dot H \vec e_{ij} - f_{vdW} \vec e_{ij} + \vec W_i
\end{split}
\label{PFD}
\end{equation}
The index $j$ is equal to $2$ for $i=1$ and to $1$ for $i=2$. Inertia in the air flow surrounding the drops is controlled by the Reynolds number of each drop, with $\vec V_i(t)$ the drop velocity,
\begin{equation}
\mathcal R_i=\frac{\rho_g V_i R_i}{\eta_g}.
\end{equation}
$\vec F_{ji}$ is the long-range aerodynamic force exerted by the drop $j$ on the drop $i$. $f_{vdW}$ is the van der Waals interaction. The correction $3/8 {\mathcal R}_i$ to the drag on each drop arises from the Oseen approximation.\cite{batchelor_introduction_2010} Consistently, the terminal velocity under gravity, noted $U^t_i$, obeys
\begin{equation}
\left(1+\frac{3 \rho_g R_i U^t_i}{8 \eta_g} \right) \; U^t_i = \frac{ 2 (\rho_\ell-\rho_g) g R_i^{2}} {9 \eta_g}.
\label{eq:terminal_velocity}
\end{equation}
For water drops in air, the associated Reynolds number ${\mathcal R}_i$ reaches $1$ around a radius $R_i\simeq 56\,{\rm \mu m}$. The term $6 \pi \eta_g a^2 \zeta'(H) \dot H \vec e_{ij}$ is the generic form of the lubrication force originating from the pressure between the two drops. $\zeta$ is a function which will be derived in section \ref{sec:lubrication_force}. The dimensionless parameter controlling the relative influence of inertia and viscous damping is the Stokes number, defined here as
\begin{equation}
\mathrm{St}\equiv \frac{\rho_\ell a (U^t_1-U^t_2)}{\eta_g}.
\end{equation}
Consider two drops of sizes in the same range, say $R_1=2R_2=3a$. In the viscous aerodynamical regime (Stokes drag), the Stokes number simplifies into: ${\rm St} = \frac 32 \mathcal G^3 $. The dimensionless number $ \mathcal G$ can therefore be interpreted as a Stokes number at the terminal velocity, to a power $1/3$. $ \mathcal G$ characterises the influence of inertia for a drop falling around the equilibrium between gravity and viscous friction. The low Stokes number regime, where inertia is negligible, is referred to as the overdamped regime. Overdamped dynamics is described by Eq.~(\ref{PFD}) without the acceleration term on the left hand side, i.e. assuming force balance at all times.

In the equations of motion, $\vec W_i$ is the thermal noise, delta-correlated in time. The noise is normalised using the fluctuation-dissipation theorem described in section~\ref{sec:diffusion}. When particles are far from each other, it leads to a relative diffusion of the two droplets with a diffusion coefficient
\begin{equation}
D=\frac{k_B T}{6 \pi\eta_g a}.
\end{equation}
The relative amplitude of aerodynamic effects and thermal diffusion is controlled by the P\'eclet number, defined as:
\begin{equation}
\mathrm{Pe} = \frac{(R_1+R_2)(U^t_1-U^t_2)}{D}
\end{equation}
The diffusive regime, where Brownian motion dominates, corresponds to the low P\'eclet number asymptotics.

\subsection{Diffusive, electrostatic and inertial regimes}
Figure~\ref{fig:ChangeParadigm}(a) is adapted from the classical textbook of \citeauthor{pruppacher_microphysics_2010}\citep{pruppacher_microphysics_2010}. It shows the collisional kernel $K$ between drops of size $R$ with one drop of size $\SI{1}{\mu m}$. Multiplied by the number of drops of radius $R$ per unit volume, $K$ gives the collision frequency of a $\SI{1}{\mu m}$ drop with drops of size $R$. The figure shows a gentle cross-over between Brownian coagulation and gravitational coagulation for drops around $R\simeq \SI{2}{\mu m}$. Figure~\ref{fig:ChangeParadigm}(b) presents our results for the same problem. The dotted-dashed blue line shows the results obtained when taking into account gravity and hydrodynamics only. Below $R=\SI{10}{\mu m}$, the kernel is ten times smaller than that in panel (a) but above $R=\SI{10}{\mu m}$, it increases much faster. The dashed red line takes into account van der Waals interaction between drops. Finally, the full model, including Brownian motion is shown in solid green line. The diffusive regime, for $R<\SI{1}{\mu m}$ is similar to that in panel (a). However, in between $R=\SI{1}{\mu m}$ and $R=\SI{10}{\mu m}$, the dominant effect turns out to be van der Waals forces.

As mentioned above, $K$ must be multiplied by the number density of drops of size $R$ to obtain the collision frequency with drops of size $\SI{1}{\mu m}$. As the density of drops generally decays rapidly with $R$, Fig.~\ref{fig:ChangeParadigm} must be interpreted with caution. The quasi-plateau in the purely diffusive regime correponds, once weighted by the density, to a decrease of Brownian coagulation rate with $R$.

In Fig.~\ref{fig:WhichMechanism}, the contributions of the dynamical mechanisms to the collision rate is analysed as a function of the two key dimensionless numbers: the Stokes number and the P\'eclet number. The ratio $\Gamma$ of the sizes of the large and small drops is kept constant. This measures the relative change of collision rate when one dynamical mechanism is suppressed. The dotted-dashed blue line is obtained using overdamped equations (no inertia). It shows that above a Stokes number ${\rm St}$ on the order of $10$ inertia is dominant. Suppressing it completely changes the collision rate. Similarly, the solid green curve is obtained by suppressing the thermal noise from the equations of motion and shows that below a P\'eclet number $\mathrm{Pe}$ of unity, Brownian coagulation is dominant. Over three decades in drop size $a$, both inertia (${\rm St}<1$) and thermal noise ($\mathrm{Pe}>1$) are inefficient, so that a third mechanism becomes dominant: electrostatic interactions. One observes that removing the van der Waals forces changes the collision rate by $50\%$ (dashed red line). The gap between the diffusive and inertial regimes constitutes the central result of this paper.

\subsection{Experimental data}
Figure~\ref{fig:comparisonexp} compares the collision efficiencies modeled here to experimental data from the literature. Experimental efficiencies roughly collapse on a master curve when plotted as a function of the Stokes number $\mathrm{St} = \rho_\ell a (U_1^t-U_2^t)/\eta_g$. They show a drop of the efficiency when inertia becomes comparable to aerodynamic effects, as predicted here and in most numerical works since Langmuir.\cite{langmuir_production_1948} The efficiency decreases when the smaller drop does not have enough inertia to cross the streamlines around the larger drop. Efficiencies for $\Gamma$ close to $1$ are less reliable and do not follow this trend, as the small velocity difference means that the drops interact for very long times that may not be reached experimentally. The most accurate experiment is that by \citeauthor{vohl_collision_2007}\cite{vohl_collision_2007}, who used a single collector drop in a controlled airflow at its terminal velocity. The insert of Fig.~\ref{fig:comparisonexp} shows the Stokes number for which $E=0.2$ (red dotted line) as a function of $\Gamma$. These measurements overlap with our computations for $\Gamma <10$. For larger size ratios, the experimental data rather follows the critical Stokes number for head-on collisions given by Eq.~(\ref{eq:critical_stokes}). Unfortunately, efficiencies around the minimum for $\mathrm{St}=1$ (ie. $6\;{\rm \mu m}$) have never been measured experimentally, nor below this cross-over value. This presents its own experimental challenges, as the critical impact parameter near the minimum is at nanometre scale. Further experimental work is needed to understand the fine details of the collision in the parameter range for which the collision frequency drops, the regime most relevant to cloud microphysics.\citep{beard_warm-rain_1993}

\section{Inertial regime}
\label{sec:inertial}
We first revisit the inertial regime, neglecting both electrostatic interactions and Brownian motion. This section is therefore devoted to aerodynamical effects and drop inertia.

\subsection{Long-range aerodynamic interactions}
In first approximation, $\vec F_{ji}$ can be deduced from the effective velocity induced by the drop $j$ at the location of the drop $i$ considered. The drag force can be linearised with respect to the velocity difference between the drop and the gas. Denoting by $\vec u(\vec r)$ the velocity field induced by the drop $j$ and taking into account the Faxén correction to the drag force,\citep{guazzelli_physical_2012} we get
\begin{equation}
\vec F_{ji}=6\pi \eta_g R_i \left(1+\frac 34 {\mathcal R}_i \right) \left(\vec u+\frac 16 R_i^2 {\vec \nabla}^2 \vec u\right)_{\vec r_i}.
\end{equation}
The Oseen solution is obtained by linearising the equations around the mean flow. The polar coordinate system is centered on the drop $j$ inducing the field. $\theta=0$ is the direction of the velocity vector $\vec V_j$. The radial velocity $u_r$ and the tangential velocity $u_\theta$ are given by
\begin{equation}
u_{r} =\frac{1}{r^{2} \sin \theta} \frac{\partial \Psi}{\partial \theta} \quad {\rm and} \quad u_{\theta} =-\frac{1}{r \sin \theta} \frac{\partial \Psi}{\partial r},
\end{equation}
where the stream function reads\citep{lamb_uniform_1911}
\begin{equation}
\Psi=-V_j R_j^{2} \sin ^{2} \theta \frac{R_j}{4 r}+\frac{3 V_j R_j^{2}}{2{\mathcal R}_j}(1-\cos \theta)\left(1-\phi_j \right),
\end{equation}
with $$\phi_j\equiv \exp \left(-\frac{r {\mathcal R}_j}{2 R_j}(1+\cos \theta)\right).$$
The velocity field reads:
\begin{align}
\frac{u_r}{V_j} &= -\frac{R_j^{3} \cos \theta}{2 r^{3}}+\frac{3 R_j^{2}}{2 r^{2} {\mathcal R}_j}\left(1-\phi_j \right) -\frac{3 R_j(1-\cos \theta)}{4 r} \phi_j, \\
\frac{u_\theta}{V_j} &= -\frac{R_j^{3} \sin \theta}{4 r^{3}}-\frac{3 R_j \sin \theta}{4 r} \phi_j.
\end{align}
The solution presents an intermediate asymptotics which coincides with the Stokes solution, the velocity field decaying as $r^{-1}$. However, at distances much larger than $R_j/{\mathcal R}_j$, the solution decays much faster, as $r^{-2}$. The Faxén correction requires the evaluation of the Laplacian:
\begin{align}
\left. {\vec \nabla}^2 \vec u \right|_r &= -\frac{3 (2R_j+r {\mathcal R}_j) \left(r {\mathcal R}_j \sin ^2\theta+4 R_j \cos \theta\right)}{8R_j r^3} \phi_j, \\
\left. {\vec \nabla}^2 \vec u \right|_\theta &= -\frac{3 \sin \theta (4R_j^2+ r {\mathcal R}_j (2R_j+r {\mathcal R}_j)(1+\cos \theta))}{8 R_j r^3} \phi_j.
\end{align}

\subsection{Lubrication force}
\label{sec:lubrication_force}
For rigid spheres, the gap $h(r)$ between the drops can be locally expanded as:
\begin{equation}
h(r) = H+R_1+R_2-\sqrt{R_1^2-r^2}-\sqrt{R_2^2-r^2}\simeq H+ \frac{r^2}{2a}
\end{equation}
as shown in Fig.~\ref{fig:schemaforces}(b). In order to regularize the lubrication force, we first introduce the slip length, which is approximately equal to the mean free path $\bar \ell$ in a gas. Slip at the interface is taken into account using the Navier slip boundary conditions $u_r = \bar \ell d u_r/dz$ at $z = 0$ and $u_r = -\bar \ell d u_r/dz$ at $z = h$. Using cylindrical coordinates, the velocity profile, in the lubrication approximation, reads:
\begin{equation}
u_r=\frac{1}{2 \tilde \eta_g} \partial_{r} P \left(z^{2}-(z+\bar \ell) h\right)
\end{equation}
where $\tilde \eta_g$ is an effective viscosity which is equal to the viscosity $\eta_g$ at large $H/\bar \ell$, but which gets smaller in the Knudsen regime $H/\bar \ell<1$ [Fig.~\ref{fig:schemaforces}(c)]. Following \citeauthor{sundararajakumar_non-continuum_1996}\cite{sundararajakumar_non-continuum_1996}, a good approximate expression of $\tilde \eta_g$ is:
\begin{equation}
\tilde \eta_g=\frac{\eta_g}{1-\frac{2}{\pi} \ln\left(\frac{3H}{3H+\bar \ell}\right)} \simeq \frac{\eta_g}{\upsilon},
\end{equation}
with $$\upsilon=1+\frac{2}{\pi} \ln\left(1+\frac{\bar \ell}{3H}\right).$$
The continuity equation integrates into $\int_0^h u_r dz=- r \dot H /2$. Integrating a second time, one obtains the pressure field, which reads:
\begin{equation}
P=3\eta_g a \zeta''(h)\dot H,
\end{equation}
with $$\zeta''(h)=-\frac{1}{ 3 \upsilon \bar \ell} \left(\frac{1}{ h}+\frac{1}{6 \bar \ell}\log \left(\frac{h}{6 \bar \ell+h}\right) \right).$$
The lubrication force is obtained by integrating the pressure over the surface:
\begin{equation}
F\simeq \int_{0}^{\infty} 2\pi r P(r) dr=-6 \pi \eta_g a^2 \zeta'(H) \dot H,
\label{EquSansN}
\end{equation}
with $$\zeta'(H)=\frac{1}{3 \upsilon \bar \ell} \left[ \left(1+\frac{H}{6 \bar \ell}\right) \log \left(1+\frac{6 \bar \ell}{H}\right)-1\right].$$
\begin{figure*}[t!]
\centering
\includegraphics{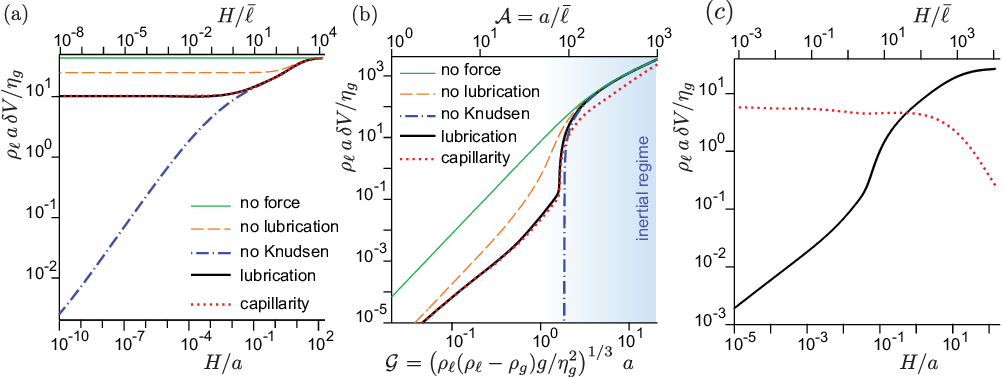}
\caption{ (a) Example of head-on trajectory in the phase space $\dot H$ vs $H$, illustrating the role of the different forces. The size ratio is $\Gamma=\frac{R_1}{R_2}=5$. $a\simeq 89.5\bar\ell$ is chosen in the critical condition for the integration performed with the regularisation by the mean free path $\bar \ell$. (b) Normal impact velocity as a function of the rescaled size $\mathcal G$ for $\Gamma = 5$. (c) Evolution of normal (black solid line) and tangential (red dotted line) velocities as a function of the gap $H$, for the critical value of the impact parameter $\delta$. $\Gamma=5$ and $\mathcal{A}=75$. }
\label{fig:headoncollisions}
\end{figure*}

A second dynamical mechanism can lead to a regularization of the lubrication force: the induction of a motion inside the liquid [Fig.~\ref{fig:schemaforces}(b)]. \citeauthor{davis_lubrication_1989}\cite{davis_lubrication_1989} have solved this problem for non-deformable drops, as a function of the fluid to gas viscosity ratio ${\mathcal N}$. The integration of the equations giving the pressure profile and the force, taking both the flow inside the drop and the mean free path into account can only be performed numerically. Here, we will make use of exact asymptotic results to derive an approximate analytical formula.

Let us consider both entrainement of the liquid inside the drop, characterized by an interfacial velocity $u_t$ and a Poiseuille contribution:
\begin{equation}
u_r=u_t+\frac{1}{2 \tilde \eta_g} \partial_{r} P \left(z^{2}-(z+\bar \ell) h\right).
\end{equation}
In the mass conservation equation, the flux now reads:
\begin{equation}
\int_0^h u_r dz=-\frac{r}{2} \dot H=u_t h-\frac{\upsilon}{12 \eta_g} h^2 (h+6 \bar \ell) \partial_{r} P.
\label{forcebal}
\end{equation}
Using the Green function formalism, the tangential stress $\sigma_t$ can be related to the tangential velocity $u_t$ by a non-local relationship. Dimensionally, one obtains the scaling law:
\begin{equation}
\sigma_t=\frac{h}{2}\partial_r P \sim {\mathcal N} \eta_g \frac{u_t}{\sqrt{ah}}.
\end{equation}
Considering this scaling law as a local relationship, the flow inside the drop would lead to a term $\sim \sqrt{ah}/{\mathcal N}$ added to $h+6 \bar \ell$ in Eq.~(\ref{forcebal}). There are therefore two possible regularisation processes. Slip occurs in the Knudsen regime, below a gap $h\sim \bar \ell$. The cross-over between a dissipation taking place in the lubrication gaseous film and in the drop takes place at $h\sim a/{\mathcal N}^2$.

We therefore propose to modify the function $\zeta'(H)$ giving the force $F$ into:
%
\begin{equation}
\begin{split}
 &\zeta'(H)=\frac{1}{3 \upsilon \bar \ell} \\
 &\times \left[ \left(1+\frac{H+s \sqrt{aH}/{\mathcal N}}{6 \bar \ell}\right) \log \left(1+\frac{6 \bar \ell}{H+s \sqrt{aH}/{\mathcal N}}\right)-1\right].
\end{split}
\end{equation}
%
where $s$ is a constant. When ${\mathcal N}$ goes to infinity, one recovers Eq.~(\ref{EquSansN}). Letting $a$ go to infinity, one gets the intermediate asymptotics associated with a drop dominated dissipation: $\zeta'(H)={\mathcal N}/s\sqrt{aH}$. Identifying with the result obtained by \cite{davis_lubrication_1989} in this limit, we find: $s=1.143... \simeq 8/7$.

\begin{figure*}[t!]
\centering
\includegraphics{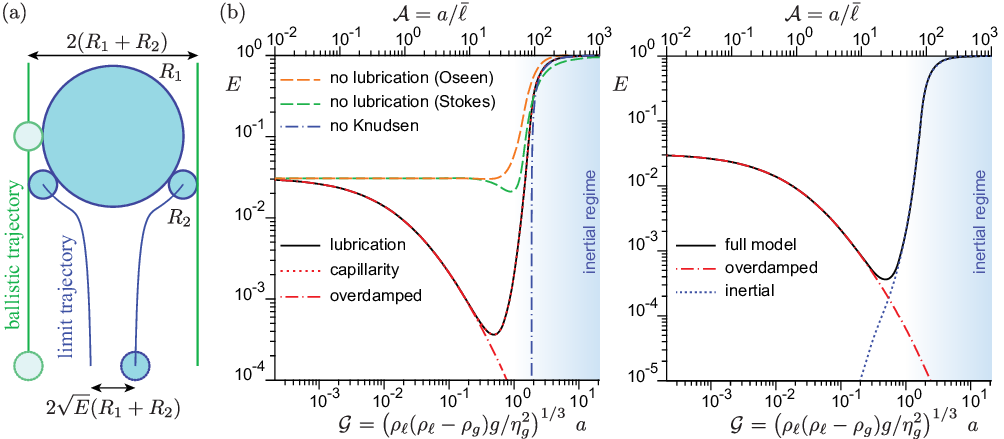}
\caption{(a) The collision efficiency $E$ is the ration between the geometric collision cross-section and the aerodynamic cross-section. Blue: trajectory for the critical impact parameter $\delta_c = \sqrt{E}(R_1+R_2)$, below which the drops always collide and above which they never do. (b) Collision efficiency for $\Gamma = 5$ as a function of the dimensionless effective radius $a/\bar \ell$ taking into account different mechanisms. Red dotted line: full solution: inertial dynamics, long-range Oseen interaction, lubrication with drop shear flow, Knudsen effects and capillarity. Dot-dashed: overdamped dynamics. Orange dashed: long-range Stokes interaction, no lubrication. Green dashed: long-range Oseen interaction, no lubrication. Blue dot dashed: full solution at vanishing mean free path.
 (c) Decomposition of the collision efficiency (black curve) as the sum of the efficiency computed in the overdamped limit (red dot-dashed curve) plus a remainder, reflecting the inertial limit (blue curve).
\label{fig:plotcollisionefficiency}
}
\end{figure*}

\subsection{Capillary-limited drop deformation}
The deformability of the drop is controlled by the liquid-vapor surface tension $\gamma$. The thermal waves have a negligible amplitude $\sim \sqrt{k_BT/\gamma}\simeq 0.2 ~{\rm nm}$ so that the relevant deformations can be predicted using hydrodynamics. The drop $j$ flattens over an extension $\delta = \sqrt{2 \tilde ah}$ which modifies the curvature $a^{-1}$ into $\tilde a^{-1}$ and the gap between drops $H$ into $\tilde H$. We denote by $u_j(r)$ the disturbance to the interfacial profile of drop $j$. The volume $\sim \delta^2 u_j$ is assumed small enough not to change the outer radius $R_j$ through the conservation of volume: the inside pressure for drop $j$ remains $2\gamma/R_j$. The gap $h(r)$ is therefore given by [Fig.~\ref{fig:schemaforces}(d)]:
\begin{equation}
h(r) \simeq H+u_1(r)+u_2(r) \simeq \tilde H+ \frac{r^2}{2\tilde a},
\label{eq:deformation_capillarity}
\end{equation}
with $$\tilde H=H+u_1(0)+u_2(0).$$
The pressure inside the lubrication film reads $3\eta_g \tilde a \zeta''(h)\dot H$. Inside the drop $j$, the pressure gradient balances the inertial term associated with the acceleration of the drop. The reference pressure is controlled by the Laplace pressure for the spherical drop, $2\gamma/R_j$. The Laplace equation evaluated at $r=0$ gives the modified curvature:
\begin{equation}
\frac{1}{\tilde a}=\frac{1}{a}-3 \tilde a \zeta''(\tilde H) \frac{\eta_g \dot H}{\gamma}.
\end{equation} 
It involves the capillary number ${\rm Ca}=\eta_g\; \dot H/\gamma$. Integrating once the Laplace equation, one obtains:
\begin{equation}
\begin{split}
&r \left(1-\frac{r^2}{R_j^2}\right)^{1/2} \gamma \frac{du_j}{dr} \\ 
&\simeq 3 \tilde a^2 \eta_g \dot H \left(\zeta'(\tilde H)\left(1-\frac{r^2}{R_j^2}\right)^{3/2}-\zeta'(h)\right).
\end{split}
\end{equation}
To obtain $u_j(0)$, one needs to integrate once more this equation, assuming that $u_j$ vanishes far from the contact zone. The term $\zeta'(h)$ leads to a logarithmic term of the form $\zeta'(H)\ln(r)$, which balances the divergence of the first term. In the outer asymptotic $\zeta'(h)\sim 1/h$, the integration can be performed explicitly, leading to $\zeta'(\tilde H)\ln(r/\sqrt{2\tilde ah})$. Using this approximation, one obtains:
\begin{equation}
\tilde H\simeq H-3 \left(\ln\left(\frac{R_1R_2}{2 \tilde a \tilde H}\right)-1\right) \tilde a^2 \frac{\eta_g}{\gamma} \zeta'(\tilde H) \dot H .
\label{eq:implicit_tildez}
\end{equation}
Equation~(\ref{eq:implicit_tildez}) is an implicit equation for $\tilde H$. It must be solved alongside Eq.~(\ref{PFD}) in which $H$ is replaced by $\tilde H$, yielding at the same time a modified drop separation and a small drop deformation given by Eq.~(\ref{eq:deformation_capillarity}).
\begin{figure*}
\centering
\includegraphics{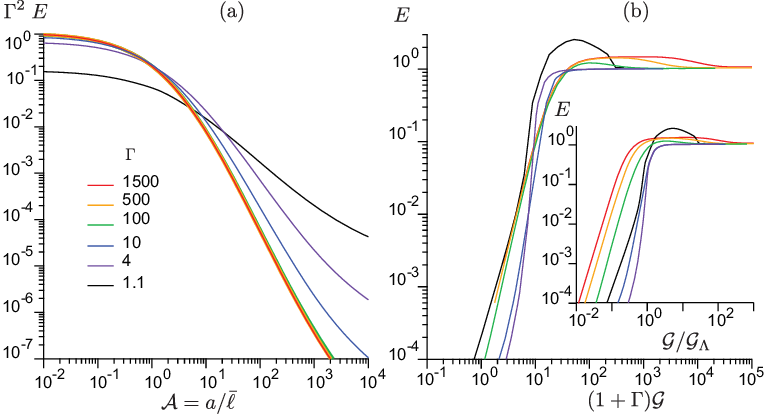}
\caption{Decomposition of the collision efficiency $E$. (a) Overdamped contribution: $\Gamma^2 E$ as a function of $\mathcal A$, for six values of $\Gamma$: $1.1$, $4$, $10$, $100$, $500$, $1500$. At asymptotically large $\Gamma$, the curves tend to a master curve. (b) Inertial contribution, defined as the difference between the efficiency and its overdamped contribution.
\label{fig:asymptoticsefficiency}
}
\end{figure*}

\subsection{Influence of the different forces}
The equations of motion governing the relative position of the two drops are integrated numerically using a Runge-Kutta scheme of order $4$, with an adaptative time step. The equations are made dimensionless using $a=1$, $\eta_g=1$ and $\rho_\ell=1$. The drops are initially at their terminal velocity, at a distance large enough to obtain results insensitive to this initial condition. In the overdamped limit, we explicitly set $d \vec V_i/dt = \vec 0$ in the governing Eq.~(\ref{PFD}) and integrate the resulting first-order coupled differential equations using also a Runge-Kutta scheme of order $4$. Over the range of parameters where the inertial and overdamped equations can both be integrated accurately, their results are identical when the drops are small enough.

To investigate the effect of the different forces, we first consider head-on collisions. Figure \ref{fig:headoncollisions}(a) shows trajectories in the phase space $(H, \dot H)$. The solid green line shows the reference case, where all aerodynamic forces are neglected; the relative velocity then remains equal to $U^t_1-U^t_2$. The dotted blue line shows the result of a calculation ignoring the regularization of the lubrication force by the mean free path. Conditions are chosen to highlight the existence of a size $a$ for which the two drops collide ($H=0$) at vanishing velocity ($\dot H=0$). When the lubrication force is removed altogether (dashed orange line), one can observe the effect of long-range aerodynamic interaction at large distances, which tends to lower the impact velocity. Lubrication forces are dominant at short separations as compared to $a$ and greatly lower the impact velocity in the absence of Knudsen effects (dot-dashed blue line). When introducing a finite mean free path $\bar \ell$ (solid black line), the lubrication force is more efficiently regularized at $H<\bar \ell$, leading to a larger impact velocity. Adding capillary deformations of the drops (dotted red line) has a very small effect on the results, in this regime.

Figure \ref{fig:headoncollisions}(b) shows the (normal) velocity at the time of impact as a function of the inertial parameter $\mathcal G$. The impact velocity is always nonzero except when the flow inside both drops is taken into account, but not the finite mean free path $\bar \ell$ (dotted blue line). In this case ($\bar \ell=0$), the impact velocity vanishes at a critical value of $\mathcal G$ below which there is no collision. This critical point will be further discussed below. When $\bar \ell$ is taken into account, one observes two regimes: at $\mathcal G>1$, the effect of $\bar \ell$ is small and the curve (black line) remains close to the curve presenting a critical point ($\bar \ell=0$). The effect of capillarity is large for big drops and tends to reduce the impact velocity. It becomes negligible at small $\mathcal G$, both because the impact velocity is small and because the drops are less deformable.

\subsection{Collision efficiency}
\label{sec:collision_efficiency}
When the drops are far from each others, they follow a linear trajectory. We define the impact parameter $\delta$ as the horizontal distance between these vertical lines. Figure~\ref{fig:headoncollisions}(c) shows the normal and tangential velocity as a function of the gap $H/a$ for particular values of $\mathcal A$ and $\Gamma$, at the critical impact parameter $\delta_c$. As $\delta$ goes to $0$, one recovers a head-on collision, for which the drops travel in straight lines with a non-zero normal velocity at impact. As $\delta$ increases, the normal velocity decreases and crosses $0$ at this critical impact parameter $\delta_c$. Figure~\ref{fig:plotcollisionefficiency}(a) shows a critical trajectory, defined by $\delta=\delta_c$, in the frame of reference of the large drop. By definition, the trajectory is tangent at the collision point ($\dot H=0$ when $H=0$). Comparing this critical trajectory to the ballistic trajectory [Fig.~\ref{fig:plotcollisionefficiency}(a)] allows one to define the collision efficiency as:
\begin{equation}
E = \frac{\delta_c^2}{(R_1+R_2)^2}.
\end{equation}
Spherical hard particles which do not interact in their trajectory have an efficiency $E = 1$, by definition. In practice, the limit trajectory is found numerically by bracketing over $\delta_c$ until $\dot H$ vanishes. The initial distance between the drops is chosen large enough to ensure that the results become independent of the choice made --~in practice, the required initial distance is around $10^2 R_1$.
\begin{figure*}
\centering
\includegraphics{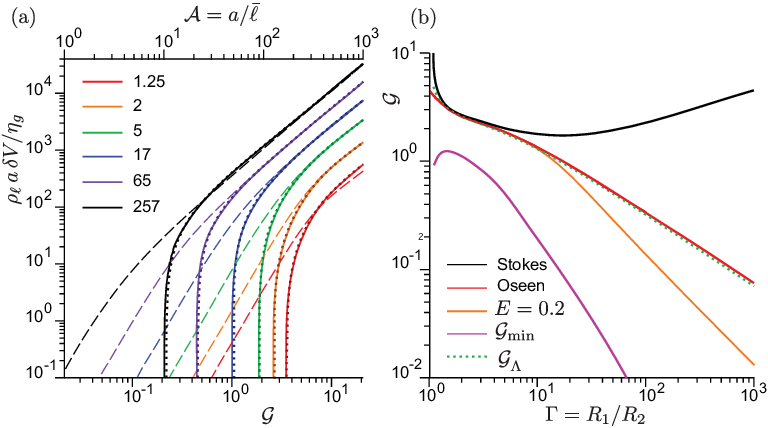}
\caption{Head-on collisions and noncontinuum effects. (a) Normal impact velocity without noncontinuum effects as a function of the rescaled drop size $\mathcal G$. The radius ratio is $\Gamma=\frac{R_1}{R_2}=5$. Solid lines: impact velocity in the limit $\bar\ell \to 0$, i.e. without noncontinuum effects. Dashed lines: initial velocity. Dotted lines: difference between the impact velocity in the presence of long-range forces only, without any lubrication, and the analytical criterion~(\ref{eq:critical_stokes}), with $\Lambda = 4.7$. (b) Critical value of $\mathcal G$ below which the impact velocity vanishes before collision, for different long-range forces. Solid black line: Stokes long-range force (with $\bar\ell \to 0$). Solid red line: Oseen long-range force (with $\bar\ell \to 0$). Dotted green line: criterion of Eq.~(\ref{eq:critical_stokes}) with $\Lambda = 17.4$. Solid orange line: isocurve of efficiency $E = 0.2$, including noncontinuum effects. Solid purple line: efficiency minimum $\mathcal G_{\mathrm{min}}$, including noncontinuum effects.}
\label{fig:headonnoncontinuum}
\end{figure*}

Figure~\ref{fig:plotcollisionefficiency}(b) shows the collision efficiency $E$ as a function of the rescaled drop size $\mathcal A$, for $\Gamma = 5$. In all cases, the ballistic limit $E = 1$ is recovered for purely inertial drops ($\mathcal G \gg 1$). Ignoring Knudsen effects but taking into account lubrication regularized by flow inside the drops ($\bar \ell = 0$; dash-dotted blue curve), $E$ vanishes below a critical value of $\mathcal G$. At this rescaled scale $a$, the critical impact parameter $\delta_c$ vanishes, which corresponds to the critical head-on collision shown in Fig.~\ref{fig:headoncollisions}. The curve provides a good approximation of the full model (solid black line and dotted red line) above $E\simeq 0.5$ and therefore captures the cross-over value of $\mathcal G$ below which the efficiency $E$ drops. The efficiency obtained without any lubrication force is plotted in dotted green line for Stokes long-range flow and in dashed orange line for Oseen long-range flow [Fig.~\ref{fig:plotcollisionefficiency}(b)]. Both approximations overestimate the efficiency at $\mathcal G<1$ and present a cross-over towards $E=1$ at large $\mathcal G$. Oseen flow reduces to Stokes flow above the small drop $2$ (downstream of it). On the opposite, below the large drop, labelled $1$, the flow velocity decays faster (as $r^{-2}$) for the Oseen approximation than for the Stokes approximation (as $r^{-1}$). As a consequence, the large drop $1$ repels less the small drop $2$ using the Oseen approximation so that the efficiency is higher. At small rescaled size $\mathcal G$, inertia becomes negligible, as confirmed by the overdamped curve (dash-dotted red curve). As the efficiency decreases with $\mathcal G$, the efficiency presents a minimum in the cross-over towards the inertial regime. This suggests decomposing $E$ as the sum of two contributions, as shown in Fig.~\ref{fig:plotcollisionefficiency}(c): an overdamped part, which can be accurately computed at vanishing inertia, and an inertial part, defined as the difference to the full calculation. Capillarity deformation of the drops turns out to be negligible in the whole range of parameters. This is due to the fact that, by construction, the efficiency curves are determined from very particular trajectories which have a vanishing normal velocity when colliding.

In order to understand the origin and the value of the minimum collisional efficiency, we make use of the decomposition of $E$ into the sum of an overdamped [Fig.~\ref{fig:asymptoticsefficiency}(a)] and an inertial [Fig.~\ref{fig:asymptoticsefficiency}(b)] contribution to the efficiency. In the limit $\Gamma \gg 1$, the smallest drop follows the streamlines around the large drop moving at its terminal velocity. The stream function $\psi(r, \theta) \simeq - U_1^t r^2 \sin^2(\theta)(1/2-3R_1/4r+R_1^3/4r^3)$ is therefore approximately constant all along the trajectory, if it is small enough to be in the overdamped regime. Initially the drops are at large distance $r$ such that $r \sin(\theta)\sim\delta$ so that $\psi = - U_1^t \delta^2/2$. The collision happens at angle $\pi/2$ at distance $r = R_1+R_2$. Equating these two values of $\psi$ gives $E=\frac{3}{2} \Gamma^{-2}$ at asymptotically large $\Gamma$. In Fig.~\ref{fig:asymptoticsefficiency}(a), the product $\Gamma^2 E$ is therefore plotted as a function of $\mathcal A$ for different $\Gamma$. At large $\Gamma$, the curves collapse on a single master curve which tends to $1$ (and not $3/2$) in the limit of vanishing $\mathcal A$. The lubrication force acts only very close to contact so that the smaller drop eventually leaves its initial streamline. This leads to the smaller prefactor observed, with a scaling still controlled by the long-range forces.

The inertial contribution to the efficiency $E$ is shown in Fig.~\ref{fig:asymptoticsefficiency}(b). For asymptotically large $\Gamma$, the drop labelled $1$ is much bigger than the second one. A change of efficiency is expected at a change of aerodynamical regime of the large drop. As $R_1=a (1+\Gamma)$, one expects that the efficiency is asymptotically controlled by the combination of parameters $\mathcal G (1+\Gamma)$. Although far to be a perfect collapse, one observes in Fig.~\ref{fig:asymptoticsefficiency}(b) much smaller variations of the inertial contribution to the efficiency, when plotted vs $\mathcal G (1+\Gamma)$. This inertial contribution presents similarities with the efficiency obtained in the limit $\bar \ell \to 0$, which vanishes below a threshold. It is therefore interesting to investigate the origin of this threshold. Figure~\ref{fig:headonnoncontinuum}(a) presents initial (dotted lines) and collisional velocities (solid lines) for head-on collisions for different $\Gamma$. The impact velocity, above the threshold, remains close to the velocity difference $U_1^t-U_2^t$. The lubrication term controls the dynamics of drops immediately before collision. Let us consider the head-on collision of two drops of mass $m_1$ and $m_2$ and let us neglect gravity and the long-range aerodynamic force. There is no inertia in the gas so that the force on the drops obeys the reciprocal action principle. As a consequence, the two body problem reduces to a single body problem with an effective mass $m_1m_2/(m_1+m_2)=\frac{4}{3}\pi \rho_\ell \mu a^3$ (with $\mu = (R_1+R_2)^3/(R_1^3+R_2^3)$) with the full force. The dynamical equation reads:
\begin{equation}
\frac{4}{3}\pi \rho_\ell \mu a^3 \ddot H= -6 \pi \eta_g a^2 \zeta'(H) \dot H.
\end{equation}
This equation integrates into
\begin{equation}
\frac{\rho_\ell a \Delta \dot H}{\eta_g}= - \frac{9}{2\mu } \Delta \zeta(H),
\end{equation}
where $\Delta$ stands for the variation between the initial and final states considered, and $\zeta(H)$ is the antiderivative of $\zeta'(H)$.

As a consequence, the lubrication film prevents the collision of drops with an insufficient initial velocity difference. The precise criterion in the absence of long-range forces is a threshold Stokes number
\begin{equation}
\label{eq:critical_stokes}
{\rm St}_\Lambda\equiv \frac{\rho_\ell a (U^t_1-U^t_2)}{\eta_g}=\frac{9}{2\mu}\Lambda, 
\end{equation}
where $\Lambda = \int \zeta'(H)\mathrm d H$ is a logarithmic factor originating from the fact that, at large $H$, $\zeta'(H) \sim H^{-1}$. $\Lambda$ depends on the dynamical mechanism regularising the lubrication pressure. Figure~\ref{fig:headonnoncontinuum}(a) shows in dotted lines the impact velocity obtained by subtracting Eq.~(\ref{eq:critical_stokes}) from the impact velocity computed only with long-range interactions, in the absence of lubrication forces. The good agreement with the full calculation, with long-range forces, using a constant $\Lambda = 4.7$ shows that there is scale separation between the near-contact lubrication and the long-range interactions.

In the viscous drag regime in which Eq.~(\ref{eq:terminal_velocity}) reduces to $U_i^t = 2(\rho_\ell - \rho_g)gR_i^2/9\eta_g$, the threshold given by Eq.~(\ref{eq:critical_stokes}) can be expressed as $ \mathcal{G}= \mathcal{G}_\Lambda$, with
\begin{equation}
\label{eq:velocity_difference_stokes}
 \mathcal{G}_\Lambda=\left(\frac{81}{4} \Lambda \frac{\Gamma^2 (1+(\Gamma-1) \Gamma)}{(\Gamma-1) (1+\Gamma)^5}\right)^{1/3}
\end{equation}
Surprisingly, this scaling is close to the scaling obtained with Oseen long-range forces [solid red line in Fig.~\ref{fig:headonnoncontinuum}(b)]. Stokes long-range forces (solid black line) displays completely different behavior as the $r^{-1}$ interaction makes the impact velocity much smaller than the terminal velocity difference over the whole range of $\mathcal G$.

This sheds light on the behavior of the efficiency near its fast inertial decrease and minimum. The values of $(\Gamma, \mathcal G)$ for which $E = 0.2$ are shown as the orange line in Fig.~\ref{fig:headonnoncontinuum}(b). For that iso-efficiency curve, $\mathcal G$ follows the scaling $\mathcal G \propto \Gamma^{-1}$ at large $\Gamma$. The value of the minimum of $E$, $\mathcal G_\mathrm{min}$, decreases faster as $\Gamma^{-2}$.

\section{Electrostatic regime}
\label{sec:electrostatic}
\subsection{Electrostatic interactions: Van der Waals and Coulombian forces}
The average droplet charge for weakly electrified clouds is\citep{harrison_microphysical_2015} $\sim e$. When the drops are far apart, they interact as two point charges $q_1$ and $q_2$ by the electrostatic potential energy $q_1 q_2/4\pi\varepsilon r$, where $\varepsilon$ is the air permittivity. The electric force can either be repulsive or attractive depending on the relative charges. In the case of clouds, one rather expects all drops to present the same sign, hence leading to a repulsion. The electrostatic force $F_e$ becomes comparable with the lubrication force $\propto \eta_g a^2 U^t/h$ driven by $U^t = 2/9 \rho_\ell g a^2/\eta_g$ at separations $h_e \sim q^2/(4\pi \varepsilon \rho_\ell g a^4)$. $h_e$ is comparable to the mean free path $\bar \ell$ for $a = 0.5\;{\rm \mu m}$ and decreases very rapidly with $a$. In disturbed weather, storm clouds develop large electric fields, the bulk of the cloud becomes negatively charged and drops carry much larger charges, up to\citep{takahashi_measurement_1973} $10^5\;e$. Here, we restrict our analysis to electroneutral drops, which is a good approximation in the bulk of warm clouds.\citep{pruppacher_microphysics_2010}
\begin{figure*}
\centering
\includegraphics{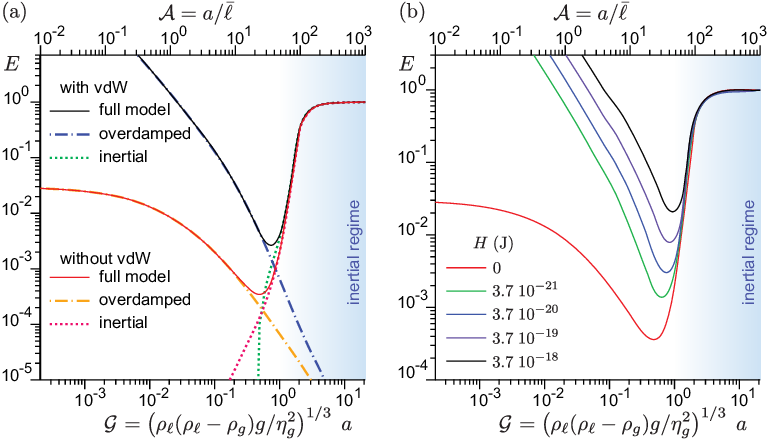}
\caption{(a) Decomposition of the collision efficiency (black curve) as the sum of the efficiency computed in the overdamped limit (red dot-dashed curve) plus a remainder, reflecting the inertial limit (blue curve), with and without van der Waals interactions. (b) Collision efficiency $E$ for $\Gamma = 5$ as a function of the dimensionless effective radius $a/\bar \ell$ for different values of the Hamacker constant: $H=0$ (red line), $H=3.7\,10^{-21}\;{\rm J}$ (green line), $H=3.7\,10^{-20}\;{\rm J}$ (blue line), which is the value for water, $H=3.7\,10^{-19}\;{\rm J}$ (purple line), $H=3.7\,10^{-18}\;{\rm J}$ (black line).
\label{fig:VdW}
}
\end{figure*}

Van der Waals forces between drops can be computed using the unretarded van der Waals pair potential. At very small separation, the disjoining pressure $\Pi(h)$ is given by the phenomenological expression
\begin{equation}
\Pi(H)=-\frac{4 \gamma \varsigma^2}{(\varsigma+H)^3}.
\end{equation}
This formula obeys the integral relation giving the surface tension: $\int_0^\infty \Pi(h) dh=-2\gamma$. Moreover, at intermediate $H$ large with respect to the molecular scale, but small with respect to $a$, one recovers the decay as $\Pi(H) \sim -4 \gamma \varsigma^2/H^3$. The force is given by integrating the disjoining pressure over the surface:
\begin{equation}
f^{\rm vdW}=\int_{0}^{\infty} 2\pi r \Pi(h) dr=-4\pi \gamma a \frac{\varsigma^2}{(\varsigma+H)^2}.
\end{equation}
This expression holds at gap $H$ comparable to $\varsigma$. \citeauthor{hamaker_london-van_1937}\cite{hamaker_london-van_1937} showed that at all separations,
\begin{widetext}
\begin{equation}
\label{eq:vdwforce}
\vec f_{ji}^{vdW}= -H \frac{32 R_1^3 R_2^3 (R_1+R_2+H)}{3(2 R_1+H)^2 (2 R_2+H)^2 (2 (R_1+R_2)+H)^2 (H+\varsigma)^2} \vec e_{ij},
\end{equation}
\end{widetext}
where $H=3.7\,10^{-20}\;{\rm J}$ is the Hamaker constant. It matches with Eq.~(\ref{eq:vdwforce}) in the small $H$ limit, provided the cut-off (molecular) length is rewritten as $\varsigma=\sqrt{H/(24\pi \gamma)}$. Note that the molecular cut-off length $\varsigma$ is equal to $0$ in the works of \citeauthor{hamaker_london-van_1937}\cite{hamaker_london-van_1937}. At large distances, the van der Waals force asymptotically tends to
\begin{equation}
f^{\rm vdW} = -H\frac{32 R_1^3 R_2^3 }{3H^7}.
\end{equation}

Figure \ref{fig:VdW}(a) compares the efficiency curves obtained with and without the van der Waals interaction forces. To a first approximation the curves are superimposable in the inertial zone but, in the overdamped regime, the attractive interactions lead to a significantly higher collision efficiency than without. The decrease in efficiency at small size $a$ is mainly due to the lubrication layer, but the residual efficiency for near-frontal collisions is significantly affected by the subdominant van der Waals interactions. In particular, the efficiency minimum is shifted by a factor $2$ in size $a$, and is almost $10$ times larger when the attractive intermolecular interactions are taken into account.

Figure~\ref{fig:VdW}(b) compares the efficiency curves obtained for different values of the Hamacker constant. It can be seen that the inertial component of the efficiency remains practically unchanged. The more attractive the interactions are, the more efficient the collisions are in the overdamped regime. It can be observed that the curves obtained are parallel: the efficiency presents an asymptotic behaviour with the scaling law $E \propto \mathcal T^{0.8}\mathcal A^{-1.7} \Gamma^{-1}$.

\begin{figure}
\centering
\includegraphics{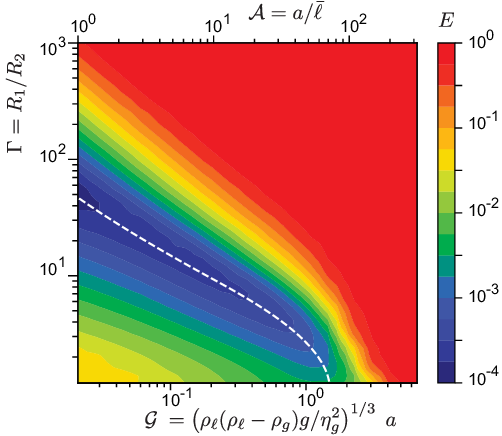}
\caption{Isocontours of the collision efficiency $E$ as a function of $\Gamma$ and $\mathcal A$. Dashed line: minimum value of the collision efficiency for at a given $\Gamma$. All aerodynamical effects are taken into account. Parameters are those of water droplets, without van der Waals interactions. For large sizes and aspect ratios, geometric behavior is recovered. Below a certain Stokes number $\mathcal G$, the efficiency decreases very rapidly and reaches a minimum.
\label{fig:E_isocontours_hydro}
}
\end{figure}

\begin{figure}
\centering
\includegraphics{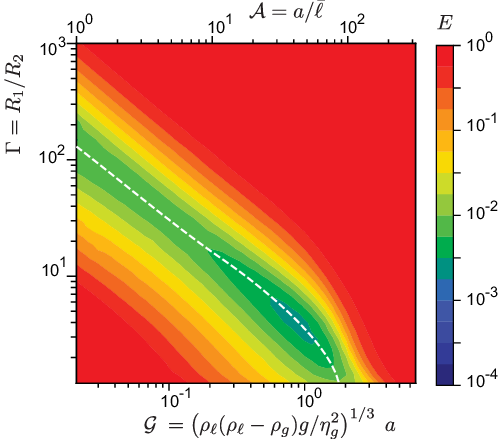}
\caption{Isocontours of the collision efficiency $E$ as a function of $\Gamma$ and $\mathcal A$. Dashed line: minimum value of the collision efficiency for at a given $\Gamma$. All aerodynamical effects are taken into account. Parameters are those of water droplets, with van der Waals interactions. The efficiency rapidly decreases from the geometric value when inertia is low enough, in the same way as the purely hydrodynamic case. van der Waals interactions increase the minimum efficiency by a factor $10$. Collisions in the overdamped regime are also more effiecient, leading to a narrower valley of collision slowdown.
\label{fig:E_isocontours_vdw}
}
\end{figure}

Figure~\ref{fig:E_isocontours_hydro} shows the efficiency $E$ as a function of the radius ratio $\Gamma$ and the rescaled size $\mathcal A$, for water drops in air, on Earth. As $\Gamma$ increases, for very large drops collecting small ones, the minimum of the efficiency decreases as well as the value of $\mathcal A$ at which it is realized (dashed line). Without van der Waals interactions, the minimum efficiency is less and less pronounced as the size $a$ increases. By contrast, taking the van der Waals interactions into account [Fig.~\ref{fig:E_isocontours_vdw}], the minimum efficiency is (roughly ten times) larger, but presents a much weaker dependance on the size $a$.

\section{Diffusive regime}
\label{sec:diffusion}

In the previous two sections, we have studied the mechanisms that control the collision efficiency in the athermal limit. In the overdamped, small size regime, we have shown that the dynamics is dominated by the lubrication layer between the drops, with attraction by van der Waals forces and long-range aerodynamic interaction playing a subdominant role. In this regime, another mechanism can play a very important role: Brownian motion. In the theoretical description of droplet aggregation processes, it is generally assumed that the collision rates induced by Brownian motion and those generated by gravity are additive. Here, we will study simultaneously the gravitational and Bronian coagulation and test this additivity assumption. We will show that thermal diffusion dominates the collision rate below a transitional Péclet number of order unity, but has a subdominant influence in the region of the parameter space which is neither inertial nor diffusive, so that the collision efficiency is minimal. This implies first redefining the collision efficiency by including the effect of thermal noise and gravity simultaneously.

\subsection{Normalisation of the thermal noise}
We now take into account the thermal noise in the equation of motion (\ref{PFD}). Following \citeauthor{batchelor_brownian_1976},\cite{batchelor_brownian_1976} we consider that the separation of time scales between the time to return to thermal equilibrium and the time for the particle configuration to change is sufficient to consider the position of the particles as constant when computing the correlations of thermal noises. In the inertial regime, diffusion is negligible. Conversely, in the regime where diffusion is important, inertial effects can be neglected. As a consequence, we will use the Stokes model of long range aerodynamic interactions rather than Oseen's model. Then, the equations for the velocity component along the direction of the axis joining the two particles and those for the two perpendicular components decouple. For simplicity, we will single out one of such velocity components and introduce the equation governing the velocity fluctuation $u_i$ along the chosen axis. As thermal diffusion is an Ornstein-Uhlenbeck process, it is convenient to write the Langevin equation under the following form,\citep{risken_fokker-planck_1989} with Einstein summation over indices:
\begin{equation}
\frac{d u_i}{dt}= -S_{ij} u_j+ W_i,
\end{equation}
with $$\langle W_i(t)W_j(t') \rangle = \mathcal{W}_{ij} \delta(t-t').$$
The index $i$ and $j$ here refer to the particle number, the three components being considered separately. It is convenient to write the relaxation rate matrix $\mathbf{S}$ under the form:
\begin{equation}
S_{ij} = \frac{9 \eta_g}{2\rho_\ell R_i^3} \alpha_{ij}.
\end{equation}
For the axis joining the center of the drops the matrix $\alpha_{ij}$ reads :
\begin{eqnarray}
\alpha_{11}&=& R_1+ a^2 \zeta'(H),\quad\quad\alpha_{22}= R_2+ a^2 \zeta'(H),\\
\alpha_{12}&=&\alpha_{21}= -\frac{R_1R_2 }{2r}\;\left(3-\frac{R_1^{2} +R_2^{2} }{r^{2}}\right) - a^2 \zeta'(H).
\end{eqnarray}
In the plane perpendicular to the axis joining the center of the drops, it reads: 
\begin{eqnarray}
\alpha_{11}&=& R_1,\quad\quad\alpha_{22}= R_2,\\
\alpha_{12}&=&\alpha_{21}= -\frac{ R_1R_2 }{4r}\;\left(3+\frac{R_1^{2} +R_2^{2} }{r^{2}}\right).
\end{eqnarray}
%
The Langevin equation integrates into, summed over indices:
\begin{equation}
u_i= \int_0^t G_{ij}(t-t') W_j(t') dt',
\end{equation}
where $G_{ij}$ denotes the the matrix elements of the Green's function $\mathbf{G}$, which formally obeys, in matrix notations,
$\mathbf{G}=\exp (-\mathbf{S} t)$. Similarly, the velocity $u_i$ can be integrated formally to give the position.

The velocity correlation function reads, with Einstein summation:
\begin{equation}
\langle u_i(t)u_j(t) \rangle=\int_0^t G_{ik}(t')G_{jl}(t') dt' \mathcal{W}_{kl}.
\end{equation}
Using the generalised equipartition of energy, 
\begin{equation}
\langle u_i(t)u_j(t) \rangle_{t\to\infty}= \frac{3 k T}{4 \pi \rho_\ell R_k^3} \delta_{ik}\delta_{jk},
\end{equation}
we deduce:
\begin{equation}
\mathcal{W}_{ij}=\frac{3kT}{4 \pi \rho_\ell} \left(\frac{S_{ij}}{R_j^3}+\frac{S_{ji}}{R_i^3}\right).
\end{equation}
In practice, at each integration step of the Runge-Kutta of order 4 algorithm, noise terms obeying a series of correlation rules described by \citeauthor{ermak_numerical_1980}\cite{ermak_numerical_1980} between positions and velocities are added to the deterministic increments. The fourth order integration scheme is recovered in the limit where the thermal diffusion is negligible. Conversely, the scheme is designed to lead to the exact diffusion result, when diffusion is dominant, provided the thermal equilibration time-scale is smaller than the typical time-scale of evolution of the geometrical configuration.

\subsection{Redefining the collision efficiency}
\label{RedefiningEfficiency}
We have  previously defined the collision efficiency as the factor encoding the influence of aerodynamics and electrostatic forces on the collision frequency of particles. The reference frequency was derived in the ballistic limit, in which the two particles of radii $R_i$ settle with a differential speed $U = U^t_1-U^t_2$, and reads: $\nu = \pi (R_1+R_2)^2 n_0 U$. Considering now the effect of thermal noise, the effect of diffusion must be included in the reference collision frequency to which the real rate is compared to define $E$. Each particle diffuses with a diffusion constant $D_i = k_B T/(6\pi\eta_g R_i)$. The problem is equivalent to the advection-diffusion of particles with a diffusion coefficient $D = D_1 + D_2$. The diffusion-advection equation for the particle concentration $n$ is, in spherical coordinates,
\begin{equation}
\label{eq:diffusionadvectionpe}
\begin{split}
&-U \cos\theta \partial_r n + U \frac{\sin\theta}{r}\partial_\theta n \\
&= D\left(\frac{1}{r^2}\partial_r (r^2 \partial_r n) + \frac{1}{r^2 \sin\theta}\partial_\theta (\sin\theta \partial_\theta n)\right).
\end{split}
\end{equation}
The boundary conditions are $n(r\to \infty) = n_0$, $n(r = R_1+R_2) = 0$. The combined effects of diffusive and advective transport on droplet growth are described by the particle flux, i.e. the collision rate, at the drop surface
\begin{eqnarray}
\nu_0 &=& 2\pi D (R_1+R_2)^2 \int_0^\pi \mathrm d \theta \, \left(\frac{\partial n}{\partial r}\right)_{r = R_1+R_2} \sin\theta\nonumber\\
&=& \pi (R_1+R_2)^2 n_0 U q(\mathrm{Pe}).
\end{eqnarray}
In the purely Brownian limit, one gets $\nu = 4\pi D (R_1+R_2) n_0$ so that $q(\mathrm{Pe}) \sim 4/\mathrm{Pe}$. In the purely ballistic limit, $q(\mathrm{Pe}) \sim 1$. Equation~(\ref{eq:diffusionadvectionpe}) has been solved analytically,\cite{simons_kernel_1986} and $q(\mathrm{Pe})$ can be expressed as
\begin{equation}
q(\mathrm{Pe}) = \frac{4\pi}{\mathrm{Pe}^2}\sum_{n=0}^\infty (-1)^n (2n+1)\frac{I_{n+1/2}\left(\frac{\mathrm{Pe}}{2}\right)}{K_{n+1/2}\left(\frac{\mathrm{Pe}}{2}\right)},
\end{equation}
with $I_n$, $K_n$ the modified Bessel functions. Care must be taken when evaluating this series.\citep{sajo_evaluation_2008} Consequently, we define the collision efficiency in the presence of diffusion and all aerodynamic effects as the ratio between the collision rate $\nu$ and the reference collision rate $\pi (R_1+R_2)^2 n_0 U q(\mathrm{Pe})$:
\begin{equation}
E_d = \frac{\nu}{\pi (R_1+R_2)^2 n_0 U q(\mathrm{Pe})}.
\label{eq:def_ed}
\end{equation}
In practice, we compute the collision frequency $\nu$ using a Monte Carlo method. We uniformly sample impact parameters over a square upstream and compute the trajectory for each sample point. The reduced colllsion rate $\nu/(\pi (R_1+R_2)^2 n_0 U)$ is estimated by measuring the ratio between the surface area of impact parameters leading to a collision to the area in the geometric case $\pi (R_1+R_2)^2$.
\begin{figure*}
\centering
\includegraphics{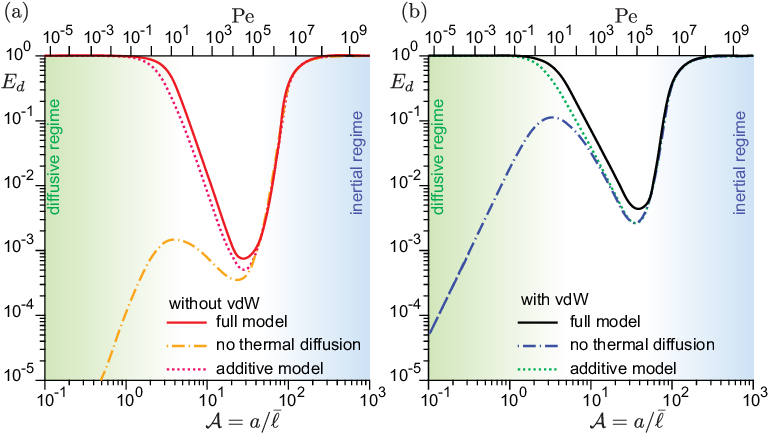}
\caption{(a) Collision efficiency $E_d$ for $\Gamma = 5$ without van der Waals interactions.  (b) Collision efficiency $E_d$ for $\Gamma = 5$ with van der Waals interactions. Solid lines: Monte-Carlo simulation of the Langevin equation, with and without van der Waals forces, taking into account thermal noise. Dash-dotted lines: previous athermal efficiency in the new definition of the diffusiogravitational efficiency $E_d$. Dotted lines: additive model for the diffusiogravitational efficiency.}
\label{fig:ThermalNoise}
\end{figure*}

\subsection{Results}

\begin{figure}
\centering
\includegraphics{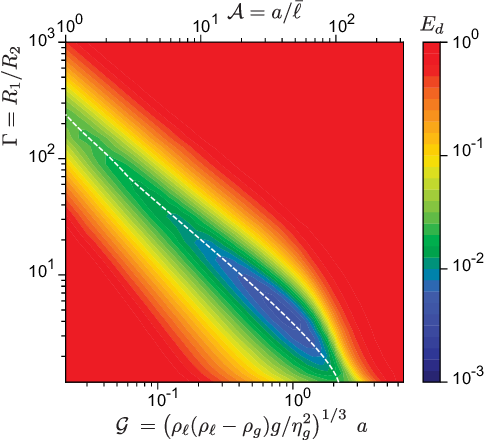}
\caption{Isocontours of the collision efficiency $E_d$ as a function of $\Gamma$ and $\mathcal A$. Dashed line: minimum value of the collision efficiency for at a given $\Gamma$. All aerodynamical effects, van der Waals interactions and thermal diffusion are taken into account. Parameters are those of water droplets. For large sizes and aspect ratios, geometric behavior is recovered; likewise, for small sizes and aspect ratios, purely diffusive behavior is recovered. Inbetween the two regimes, there is a valley of lower collision rate where electrostatic effects dominate. 
\label{fig:ed_map}
}
\end{figure}

The collision efficiency $E_d$ is shown for $\Gamma = 5$ without van der Waals forces in Fig.~\ref{fig:ThermalNoise}(a). At large inertia ($\mathcal G >1$), the efficiency curve (solid line) collapses with athermal results (dash dotted line) and there is no effect of diffusion, as expected. At small Péclet number, fully diffusive behavior ($E_d \to 1$) is recovered asymptotically. Note that, with the new definition of the efficiency taking into account diffusion, athermal collisions become vanishingly inefficient at small Péclet number. The effect of diffusion is visible at the collision efficiency minimum, even when it is reached at high Péclet numbers ($\sim 10^5$) where diffusion would be expected to be negligible at a glance. The effect of Brownian motion on particle growth is often modelled in the literature\citep{greenfield_rain_1957} by simply adding together the purely diffusive particle collision rate $4 \pi D (R_1+R_2) n_0$ with the purely athermal particle collision rate $\pi (R_1+R_2)^2 E U n_0$ computed previously. With the definition of $E_d$ (\ref{eq:def_ed}) proposed there, this leads to
\begin{equation}
E_d^\mathrm{sum} = \frac{4/\mathrm{Pe}+E}{q(\mathrm{Pe})}.
\end{equation}
Note that $q(\mathrm{Pe}) \sim 4/\mathrm{Pe}$ in the small Péclet number limit, leading to $E_d^\mathrm{sum}\to 1$ for $\mathrm{Pe}\to 0$. The resulting curve (dotted line) does not collapse over the full Monte-Carlo solution: diffusion and aerodynamic effects are not additive on particle growth. Results with van der Waals interactions are shown in figure \ref{fig:ThermalNoise}b. Similarly, the purely diffusive and purely ballistic regimes are recovered asymptotically, with diffusion present in the gap. Likewise, the additive model $E_d^\mathrm{sum}$ does not accurately reproduce the simulation results. Van der Waals interactions and diffusion enhance each other in a non-trivial way near the gap. Figure~\ref{fig:ed_map} shows the combined diffusiogravitational efficiency $E_d$ for water drops on Earth as a function of $\mathcal A$ and $\Gamma$, with van der Waals interactions. The gap in the efficiency is about $10^{-3}$, and gets slightly shallower, narrower and shifted to smaller sizes as the size ratio $\Gamma$ of the droplets increases.

\section{Concluding remarks}
In this article, we have analysed the influence of relevant dynamical mechanisms on the collision efficiency of drops suspended in a gas. The largest drops, falling under the effect of gravity, merge when their motion is inertial, meaning that their Stokes number $\mathrm{St}$ is larger than $1$. The smallest drops have a motion controlled by thermal diffusion, below a unit Péclet number. The main result of this paper is the existence of a range of drop sizes for which neither inertial nor diffusive effects are dominant, resulting in a large decrease of the collision efficiency. In this intermediate regime, it is the gaseous lubrication film separating the drops that prevents them from merging. In addition, van der Waals forces become non-negligible. The code used here allows to compute the efficiency over the whole size range in clouds, both for equally-sized drops and for drops of very different sizes.

As the outcome of a collision is binary (either the droplets merge or they do not), and the growth rate increases rapidly with the drop size, the growth process is extremely sensitive to minute details of the collision. We have neglected here added mass, the Basset/history force, shear, all rotation and torque effects,\citep{maxey_equation_1983,mordant_velocity_2000} but also non-aerodynamic effects such as retardation in the Van der Waals force,\citep{gregory_approximate_1981} thermophoresis due to temperature gradients and diffusiophoresis due to water vapor gradients.\citep{friedlander_smoke_2000} For millimetric drops, capillary deformations become important around a Weber number of $1$. Drop deformation flattens the drops, which changes their settling speed and ultimately leads to break-up.\citep{reyssat_shape_2007,villermaux_single-drop_2009} Capillary instabilities can also be triggered during the collision, leading to the formation of smaller droplets by various fragmentation processes of liquid filaments and sheets.\citep{villermaux_fragmentation_2007,testik_toward_2011,testik_first_2017}

Turbulence is often hypothesized to be the dominant mechanism through which large drops can form, leading to rain.\cite{wilkinson_caustic_2006,grabowski_growth_2013,vaillancourt_review_2000,vaillancourt_microscopic_2002,benmoshe_turbulent_2012} In clouds, the Reynolds number is around $10^7$, with a dissipation rate of $\epsilon \sim 10^{-3}-10^{-2}\SI{}{m^2.s^{-3}}$. This sets the Kolmogorov scales as $\ell_K \sim \SI{1}{mm}$ for the typical length, $u_K \sim \SI{10}{mm.s^{-1}}$ for the velocity and $\tau_K \sim \SI{0.1}{s}$ for the time.\cite{shaw_particle-turbulence_2003,mellado_cloud-top_2017} Droplets are therefore entirely in the dissipative range of scales. When the droplets have low inertia, they follow the streamlines of the local sub-Kolmogorov uniform strain field. \citeauthor{saffman_collision_1956}\cite{saffman_collision_1956} have computed the collision kernel $K$ between two droplets in this case. Assuming $E = 1$, it scales as the product of their collision cross-section $\pi (R_1+R_2)^2$ by the relative velocity, given by the velocity gradient over the separation $R_1+R_2$:
\begin{equation}
K \propto \pi (R_1+R_2)^2 \frac{\partial u(r)}{\partial r}(R_1+R_2) \propto \frac{(R_1+R_2)^3 u_K}{\ell_K}.
\end{equation}
When the droplets have finite inertia, they can leave the streamlines. At low but finite inertia, this leads to fractal clustering in regions of low vorticity, which enhances the local density and thus the collision rate set by the local shear.\cite{bec_fractal_2003,bec_clustering_2005,bec_multifractal_2005,bec_turbulent_2010,chun_clustering_2005} At higher inertia, droplets can be slung away with large accelerations by local vortices.\cite{voskuhle_prevalence_2014,falkovich_sling_2007,salazar_inertial_2012} Asymptotically, they behave like molecules in a gas, with inertia acting as temperature.\cite{abrahamson_collision_1975} $K$ scales in this case as\cite{pumir_collisional_2016}
\begin{equation}
K \propto (R_1+R_2)^2 u_K \sqrt{\mathrm{St}_K} \exp{\left(-\frac{\mathrm{St}_c}{\mathrm{St}_K}\right)},
\end{equation}
with $\mathrm{St}_K = 2 \rho_\ell (R_1^2+R_2^2)/(9\eta_g \tau_K)$ the Stokes number of the droplets in the turbulent flow and $\mathrm{St}_c$ a constant. $\mathrm{St}_K$ crosses $1$ around $\SI{40}{\mu m}$, meaning that the sling mechanism is negligible around the electostatic regime. Likewise, the terminal velocity is larger than the velocity difference $u_K (R_1+R_2)/\ell_K$ above $\SI{0.5}{\mu m}$, such that local shear, even enhanced by preferential concentration, is inefficient across the whole range of sizes in clouds. Turbulence could also induce collisions through rare events not described by these mean-field effects. Intermittency can lead to very high particle accelerations \cite{siebert_towards_2010,toschi_lagrangian_2009}, which could translate to a high collision rate. However, this would necessarily involve only a small fraction of the droplet population, which might not be enough to produce a sizeable amount of rain \cite{wilkinson_large_2016}. The mechanisms of warm rain formation still remain elusive to this day.\cite{morrison_confronting_2020}

The code to compute the efficiency, both for the athermal deterministic regime and the Monte-Carlo solution to the Langevin equation, is available online.\cite{noauthor_notitle_2024}

\bibliography{ReviewCollisionPOF}

\end{document}